# Experimental evaluation of xApp Conflict Mitigation Framework in O-RAN: Insights from Testbed deployment in OTIC

Abida Sultana*, Cezary Adamczyk[†], Mayukh Roy Chowdhury*, Adrian Kliks[†‡], Aloizio Da Silva*
* Commonwealth Cyber Initiative, Virginia Tech, USA, {abidas, mayukhrc, aloiziops}@vt.edu
[†] Poznan University of Technology, cezary.adamczyk@doctorate.put.poznan.pl, adrian.kliks@put.poznan.pl
[‡] Rimedo Labs, Poland, adrian.kliks@rimedolabs.com

*Abstract*—Conflict Mitigation (CM) in Open Radio Access Network (O-RAN) is a topic that is gaining importance as commercial O-RAN deployments become more complex. Although research on CM is already covered in terms of simulated network scenarios, it lacks validation using real-world deployment and Over The Air (OTA) Radio Frequency (RF) transmission. Our objective is to conduct the first assessment of the Conflict Mitigation Framework (CMF) for O-RAN using a real-world testbed and OTA RF transmission. This paper presents results of an experiment using a dedicated testbed built in an O-RAN Open Test and Integration Center (OTIC) to confirm the validity of one of the Conflict Resolution (CR) schemes proposed by existing research. The results show that the implemented conflict detection and resolution mechanisms allow a significant improvement in network operation stability by reducing the variability of the measured Downlink (DL) throughput by 78%.

*Index Terms*—O-RAN, Near-RT RIC, xApps, conflict mitigation

## I. INTRODUCTION

Open Radio Access Network (O-RAN) has led to a paradigm shift in legacy Radio Access Network (RAN) by introducing openness, interoperability, and intelligence. Standards introduced by O-RAN Alliance aim to enable highly automated optimization in multi-vendor network environments [1]. This includes defining a number of RAN control agents that co-exist at various levels of the network. These agents include modular applications (xApps, rApps) deployed in RAN Intelligent Controllers (RICs). Managing the complexity of deployments with such heterogeneity is one of the biggest and most important challenges that the industry is facing with regard to O-RAN.

Specifically, the multiplicity of applications deployed in RICs is likely to cause interoperability problems, where control decisions from these applications collide with each other [2]. An example of such collision is when decisions made by two applications directly contradict each other, i.e., modify the same parameter but with different values. This leads to unpredictability of the network and can cause deterioration of the Quality of Service (QoS) for end users.

To accelerate efforts to validate O-RAN related developments, the O-RAN Alliance qualifies selected laboratories as Open Test and Integration Centers (OTICs). These OTICs act as centers dedicated to testing O-RAN-compliant solutions from multiple vendors in relevant heterogeneous network environments based on real hardware and software. Experiments conducted in OTICs shall act as benchmarks for interoperability and performance of solutions for O-RAN. In addition, testing and evaluation of O-RAN applications (i.e. xApps, rApps) in an O-RAN testbed is crucial to guarantee network trustworthiness and resilience before deploying them in a production environment.

In this article, we implement a Conflict Mitigation (CM) mechanism in a testbed using real-world hardware and software with Over The Air (OTA) Radio Frequency (RF) transmission to verify its validity to detect and resolve conflicts between xApps in O-RAN deployments in a realistic test scenario. Furthermore, we evaluate the efficiency of the CM method based on measurements carried out on the Commonwealth Cyber Initiative (CCI) xG Testbed which is an OTIC in Washington DC metro area [3].

### A. Related Work

The concept of CM is not yet widely discussed in O-RAN specification, although it has gained more attention recently as the development of O-RAN unfolds. The O-RAN Alliance Work Group 3 recently published a technical report that discusses challenges and potential solutions to CM in Near-Real Time RIC (Near-RT RIC) [2]. Although it provides insight into the procedural approach to CM activities in O-RAN, it does not cover the validation of the proposed solutions in realistic deployments.

Adamczyk and Kliks [4] proposed a framework named Conflict Mitigation Framework (CMF) that allows robust CM in O-RAN. Their approach enables the detection and resolution of O-RAN-specified conflicts [5]. It requires that all E2 control messages are processed by the CM function of the Near-RT RIC before the message gets sent towards the target E2 node, but does not pose any additional requirements towards xApp logic with respect to the exchange of conflict information.

Armstrong et al. [6] proposed an extension to O-RAN Service Management and Orchestration (SMO) architecture including a dedicated logical function to predict network performance degradation due to configuration changes performed. The method is based on statistical correlation and works in the Non-RT control loop. Wadud et al. [7] described another CM solution for Near-RT RIC based on cooperative bargain game theory to resolve detected control conflicts.

Bonati et al. [8] introduced Colosseum, a large-scale O-RAN compliant channel emulator capable of modeling complex scenarios. In [9], it is used as a digital twin, while in [10], it facilitates the large-scale evaluation of xApps based on Machine Learning (ML). Unlike traditional setups, Colosseum does not utilize OTA RF transmission between the RAN and User Equipments (UEs); instead, it emulates RF channels and relies on RF cables for communication. Building upon Colosseum, Del Prever et al. [11] proposed PACIFISTA, a framework for CM that leverages the Colosseum emulator to provide pre-deployment insights for mitigating conflicts and optimizing xApp/rApp deployment. PACIFISTA employs a profiling pipeline to conduct sandbox tests, generate statistical profiles, predict and assess conflicts, and identify affected Key Performance Measurements (KPMs) and control parameters.

Abida et al. [12] demonstrated a time-based approach for detecting and resolving direct conflicts in an O-RAN setup leveraging another established O-RAN testbed, the CCI xG Testbed, which is also approved as North American OTIC [3]. It is based on Software-defined Radio (SDR) and open-source cellular stacks in addition to Commercial Off-The-Shelf (COTS) hardware. The CCI xG Testbed enables validation of disaggregated RAN components (Open Central Unit (O-CU), Open Distributed Unit (O-DU), Open Radio Unit (O-RU)) and multi-vendor integration over OTA networks, supporting interactions with RICs and microservices incorporating techniques based on Artificial Intelligence (AI) and ML.

### B. Major Contributions

As shown, the current landscape of the O-RAN Alliance specifications and related publications does not cover the concept of CM evaluation in real-world O-RAN deployments – all existing works on CM in O-RAN are based on simulations or real-world testbeds with emulated RF channels. Our goal is to perform the first evaluation of a CM method for O-RAN-compliant testbed, which adheres to all O-RAN specifications with OTA RF transmission between RAN and UEs. This allows us to evaluate the performance of our approach in a realistic environment, capturing the complexities and challenges of real-world deployments. In our previous work, we demonstrated an initial implementation of time-based conflict mitigation with single UE [12], but it lacked comprehensive performance evaluation on the testbed in a dynamic environment with multiple UEs. To this end, the following are the major contributions of this work:

1) Experimental evaluation of CMF for xApps in Near-RT RIC is performed on a 5G O-RAN platform based on SDR in the Virginia Tech CCI xG testbed [3]. To the best of our knowledge, this is the first-ever testbed evaluation of CMF and the first testbed evaluation of CM measures using OTA RF connectivity between RAN and UEs.
2) The experiment is carried out using an O-RAN network deployed with open source srsRAN gNB stack, Open5GS 5G Core, O-RAN Software Community (O-RAN SC) Near-RT RIC. The RIC runs two custom-made conflicting xApps, which allocate maximum Physical Resource Blocks (PRBs) quota to slices.
3) The CMF proposed in [4] is implemented in the Near-RT RIC, which includes both Conflict Detection (CD) and a priority-based Conflict Resolution (CR) mechanisms.
4) A performance evaluation of CMF is done through multiple test-runs by comparing a performance metric, i.e., UE-level Downlink (DL) throughput, of the network with and without the CM measures in place.

## II. BACKGROUND

### A. Near-RT RIC Functionality

For the purpose of optimizing RAN functionality, the xApps are capable of monitoring and controlling E2 nodes. The RIC Message Router (RMR) enables the required channel interoperability between xApps and other components. Furthermore, the Shared Data Layer (SDL) enables xApps and other modules to store data for analysis, RAN control, and resource management within the Near-RT RIC [5].

Near-RT RIC supported functions enable xApps to execute configurable service logic on E2 Nodes using global procedures such as E2 Setup, RIC Subscription, RIC Indication, and RIC Control. Within the E2 Node, xApps can register to four RIC Services defined by E2 Application Protocol (E2AP) specifications. **REPORT** service enables the E2 Node to send reports to Near-RT RIC with information from the E2 Node along with parameters such as the number of connected UE, DL and Uplink (UL) bitrates of the base station [13]. **INSERT** service enables the E2 Node to ask Near-RT RIC for control guidance and temporarily halt call processing. **CONTROL** service supports Near-RT RIC to start or continue an xApp-defined process in the E2 Node. Synchronous messaging is used in this service, as long as the Near-RT RIC requests acknowledgment of the E2 Node's control request. **POLICY** service enables the E2 Node to automatically carry out a particular policy when the trigger event takes place by the xApps. E2 agent forwards all RAN control messages to the connected E2 node.

### B. xApp-driven RAN control

Intelligent RAN control in O-RAN's near real-time control loop is conducted through xApps deployed in Near-RT RICs. Each xApp is designed to perform specific network optimization tasks, including traffic steering, energy efficiency, slice control, anomaly detection, and various other functions. Mobile Network Operators (MNOs) can select xApps from a wide range of third-party providers or develop them internally to ensure that network operation is optimized according to their specific operational requirements.

An example of such RAN optimization driven by multiple xApps is the combination of Traffic Steering (TS) and Slice Management (SM). For instance, a TS xApp may route the traffic in the RAN as to maintain balanced load across cells, while an SM xApp configures the resource allocation for slices to ensure adequate QoS is maintained for prioritized services. Cooperation between both xApps is required to avoid conflicts between TS and SM control decisions.

*C. Types of Conflicts*

The RAN architecture with multiple network control agents is enabled with O-RAN's RICs. As xApps from various third-party vendors may co-exist, it is likely that conflicts between xApps' reconfiguration decisions will arise. Such conflicts may cause significant degradation of network performance, leading to unpredictable behavior and a lack of trust in the network service.

O-RAN specification for Near-RT RIC [5] architecture provides the following categorization of control conflicts: **Direct** conflicts with control decisions on the same parameter (i.e., control target), **Indirect** conflicts with control decisions on parameters influencing the same network area, and **Implicit** conflicts with control decisions clashing in non-obvious ways. All of these conflicts require different approaches to detect, and implicit conflicts pose the greatest challenge. Near-RT RIC's CM function shall cover both the detection and resolution of all conflicts between the control decisions of xApps deployed in the RIC.

Conflict resolution methods include modifying and rejecting conflicting control decisions to lessen the negative impacts of control conflicts on network operation. For example, Near-RT RIC may prioritize one of the xApps, and in case of conflict between control decisions from multiple applications, the RIC may reject control decisions from all non-prioritized xApps. Another example includes modifying the parameters of conflicting control decisions to influence the network in an optimal way.

## III. METHODOLOGY

*A. System Model*

The main goal of this paper is to experimentally confirm the validity of the considered CM scheme using a testbed network infrastructure operating OTA. The experiment is carried out using a dedicated O-RAN testbed set up in the North American OTIC in Washington DC/Arlington VA. The architecture of the configured network used as a testbed consists of components compliant with O-RAN specifications.

The system diagram of the deployment is shown in Figure 1. The RAN setup consists of a single gNodeB (gNB) using open source srsGNB software from the Software Radio Systems RAN (srsRAN) project [14]. The gNB hosts two network slices (named A and B), both handling enhanced Mobile Broadband (eMBB) traffic. Both slices are accessed via a single cell on frequency 3600 MHz. Slice A aims to provide services to prioritized UEs, while slice B handles traffic from regular non-prioritized UEs.

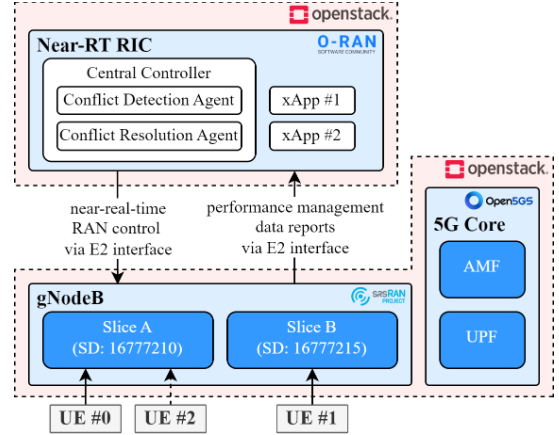

Fig. 1: End-to-end system diagram of the deployment

Network traffic is generated by up to three UEs capable of connecting to the configured gNB, with UEs #0 and #2 setup to connect to Slice A and UE #1 to Slice B.

The deployed RIC is based on the open source O-RAN SC Near-RT RIC software. It hosts 2 xApps, which optimize slice-level PRB allocations using action 6 from style 2 of CONTROL services in E2 Service Model-RAN Control (E2SM-RC) [15]. While both deployed xApps adjust slice-level PRB allocations, they follow different optimization targets. The control logic for both xApps was designed subjectively to demonstrate a practical approach toward achieving the optimization targets. While there may exist implementations closer to being optimal, the key focus is on illustrating that the logic effectively steers toward the desired outcome.

The xApp #1 aims to prioritize traffic from prioritized UEs that connect to slice A. To do so, xApp #1 allocates proportionally more of available PRBs to slice A. With total number of PRBs available to the cells $P$, total number of UEs served by gNB $U$, number of slices in gNB $S$, and number of UEs served by slice A within the gNB $U_A$, number of PRBs $p_{1,A}$ allocated to slice A is calculated by xApp #1 as shown in Equation (1) and number of PRBs allocated to non-prioritized slices other than A $p_{1,s}$ according to Equation (2). Calculations are based on priority PRB ratio $r_A$ and non-priority PRB ratio $r_s$. The ratio $r_A$ indicates the proportion of PRBs allocated based on the number of UEs connected to the prioritized slice. Meanwhile, the ratio $r_S$ represents the allocation of the remaining PRBs, after resources have been assigned to the prioritized slice, distributed equally among the other non-prioritized slices. In general, xApp #1 allocates PRBs based on the average of two factors: the ratio corresponding to the slice type ($r_A$ or $r_s$) and a factor inversely proportional to the number of slices ($1/S$). This ensures that the logic balances between slice priority ratios and overall fairness.

$$r_A = \frac{U_A}{U} \qquad p_{1,A} = P \cdot \frac{r_A + 1/S}{2} \qquad (1)$$

$$r_s = \frac{1 - r_A}{S - 1} \qquad p_{1,s} = P \cdot \frac{r_s + 1/S}{2} \qquad (2)$$

While xApp #1 prioritizes slice A, xApp #2 aims to prioritize an even split of PRBs across all network slices within gNB. The allocation of PRBs for all slices $p_2$ performed by xApp #2 is carried out as per Equation (3).

$$p_2 = P/S \qquad (3)$$

Both xApps are concurrently active in the network, with the relevant xApp logic triggered periodically every 10 seconds. After every calculation of the PRB allocations, the xApps compose and send the relevant E2 CONTROL messages to apply the calculated allocations. A Central Controller entity is implemented within the Near-RT RIC. It performs the RIC's CM function by hosting CD Agent and CR Agent components, as envisioned in CMF [4].

### B. Conflict Detection Mechanism

The control conflict between xApps in the considered test scenario is a direct conflict between xApp #1 and xApp #2. Both of these xApps modify the slice level PRB allocation but have different goals, leading to a direct conflict. The conflict between xApps is detected as defined in [4] – Near-RT RIC tracks all control decisions of xApps and validates if any of the decisions are made against the same control target. If such conflicting control decisions are found, then the information about the conflict is passed into CR Agent for resolution. This procedure is shown in Figure 2.

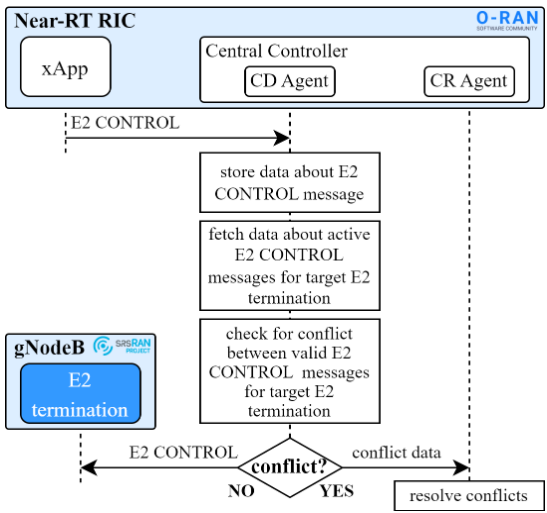

Fig. 2: Conflict Mitigation procedure

### C. Conflict Resolution Mechanism

In the test scenario, the CR approach is tailored to the specific roles of the xApps deployed in the system. xApp #1 is responsible for optimizing network performance for high-priority traffic in Slice A, making it the preferred xApp in cases of control conflicts. When both xApp #1 and xApp #2 issue E2 control messages that are active simultaneously and target the same slice, the CD Agent identifies the direct conflict. The CD Agent forwards this information to the CR Agent, which consistently resolves the conflict by rejecting the control decision from xApp #2. As a result, only the control action from xApp #1 is applied to the network, ensuring that the high priority traffic optimization is not affected by competing xApp decisions

## IV. PERFORMANCE EVALUATION

### A. Experimental Setup

The described system setup in Figure 1 was implemented in the CCI xG Testbed [3]. The Near-RT RIC provided by O-RAN SC is deployed in an open-stack Virtual Machine (VM) on CCI xG Testbed, setup with 8 GB of RAM, 8 CPU, 512 GB of storage, and running on Ubuntu 20.04 OS. The Near-RT RIC is deployed as a Docker container with the necessary interfaces to communicate with the RAN node. The xApps are also on-boarded as Docker containers inside the RIC platform, with all related sub-components and the service model compatible with the RAN environment. The srsRAN 5G gNodeB is deployed in another open-stack VM, along with Open5GS Core. For the RF function of the RAN, an Universal Software Radio Peripheral (USRP) x310 (20 MHz bandwidth) located in the Radio Ceiling Grid (indoor component of CCI xG Testbed) is used as the RF frontend, which is a 2 Tx and 2 Rx antenna capable of 2x2 Multiple Input Multiple Output (MIMO) operation. Using a 10G Ethernet cable, it connects to the VM. For the UEs, 3 Samsung phones with programmable SIM cards were used. The experimental setup of the implementation is shown in Figure 3.

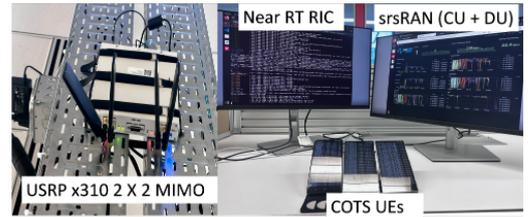

Fig. 3: Experimental Setup

### B. Evaluation methodology

The considered scenario to evaluate the efficiency of the implemented CM measures consists of two phases. First, both slices A and B are enabled, and 1 UE is connected to each slice. After a minute, a third UE connects to slice A. This causes a change in the slice-level PRB allocations computed by both xApps running in the RIC. Due to the logic implemented within the xApps, such circumstance causes a difference in the allocations of both xApps, hence leading to a direct control conflict. Two cases are considered: without any CM measures and with CM measures in place. Regardless of the case, UE-level DL throughput is observed and logged to enable the evaluation of the network performance during the execution of the experiment.

*C. Results*

To evaluate the implemented solution, twenty test runs were carried out according to the descriptions in Section IV-B: ten without and ten with CM measures (namely, CMF) enabled. The testbed setup, other than presence of CM measures or the lack thereof, was identical for all runs. Data captured for each test run was limited to one minute before the third UE connects to slice A and up to 6 minutes after. For each run, DL throughput for each UE was plotted over time, with the plots shown in Figure 4.

To ensure reliable results, the UE-level DL throughput data was processed to exclude records from the first two minutes of the test run (i.e., the considered data start one minute after the third UE connects to slice A), as it could reflect transient behaviors during network stabilization. For each UE, the average throughput and its variability (standard deviation) are calculated. To understand overall trends, the computed averages and variabilities for all UEs in a run are aggregated to derive run-level statistics. These include the mean performance across all UEs and the average variability observed. Finally, the results are further aggregated at the run category level, comparing performance across the "No CM" and "CMF" runs. The results are presented in Table I.

TABLE I: Experiment Results

| Category | Run | Downlink Throughput [Mbps] | |
|---|---|---|---|
| | | Mean | Standard Deviation |
| No CM | #1 | 14.769 | 2.622 |
| | #2 | 12.567 | 2.899 |
| | #3 | 11.972 | 2.287 |
| | #4 | 15.984 | 4.875 |
| | #5 | 14.143 | 3.913 |
| | #6 | 14.603 | 2.181 |
| | #7 | 12.673 | 2.640 |
| | #8 | 13.007 | 2.929 |
| | #9 | 14.473 | 2.559 |
| | #10 | 17.443 | 2.729 |
| | **Average** | **14.163** | **2.964** |
| With CMF | #1 | 15.117 | 0.471 |
| | #2 | 10.731 | 1.198 |
| | #3 | 17.545 | 0.985 |
| | #4 | 13.481 | 1.509 |
| | #5 | 14.400 | 1.126 |
| | #6 | 14.773 | 0.144 |
| | #7 | 14.734 | 0.109 |
| | #8 | 14.726 | 0.424 |
| | #9 | 17.579 | 0.281 |
| | #10 | 17.445 | 0.242 |
| | **Average** | **15.053** | **0.649** |

The network operation is generally stable in all runs while there are only two UEs connected. Then, in all runs without any CM measures, significant fluctuations of UE-level DL throughput can be observed for all UEs when the third UE connects to slice A (that is, after the first minute). Specifically, taking the run "No CM #4" as an example, DL throughput of UE #1 oscillates between approximately 10 and 24 MB/s. Similar disturbances are observed for other UEs, indicating unpredictable QoS for all connected devices. This is caused by the unmitigated conflict between xApps that periodically execute contradicting control decisions.

In the runs with CM enabled, when the third UE connects to the network, the DL throughput changes for each UE as the network adapts to the new PRB allocations, but then remains stable. In contrast to the runs without CM, no significant metric fluctuations are observed. This is due to the CM measure causing control decisions from xApp #1 to take precedence over decisions from xApp #2, allowing for the network to reach a stable state.

The behavior described above is reflected in the DL throughput variability as represented by Standard Deviation (SD) presented in Table I. The average SD of the DL throughput across all runs without CM measures equals 2.964 and is more than four and a half time higher than for runs with CMF enabled, where the calculated SD is 0.649.

Based on the observations described above, it can be stated that the implemented CM measures successfully neutralized the negative impacts of the direct conflict between xApps #1 and #2 by reducing DL throughput metric variability by 78%. Hence, the network operation with these measures in place is much more stable and predictable in contrast to the erratic behavior of the network without the CM measures enabled. The implemented CD and CR mechanisms do not have any negative impact on the network resources and the total throughput of the RAN.

## V. CONCLUSION

This study used real-world components deployed in CCI xG testbed in Virginia, USA, to experimentally validate the CMF. The results show that the tested CM method is successful in resolving direct conflicts between xApps deployed in the Near-RT RIC. The experiment specifically shows how to prioritize control decisions from particular xApps, such as those that optimize high-priority traffic, to maintain system stability and guarantee optimal network performance.

The presented work bridges the gap between theoretical study and real-world application by offering a strong framework for identifying and resolving conflicts in O-RAN-compliant environments. Furthermore, the testbed configuration provides a reproducible framework for validating conflict resolution and network optimization, enabling scalable and reliable O-RAN deployments in multi-vendor and commercial environments. Future works will evaluate the proposed framework in terms of latency-awareness, as well as consider xApps used for other use cases, e.g., energy efficiency optimization. We also plan to extend this towards indirect and implicit conflicts.

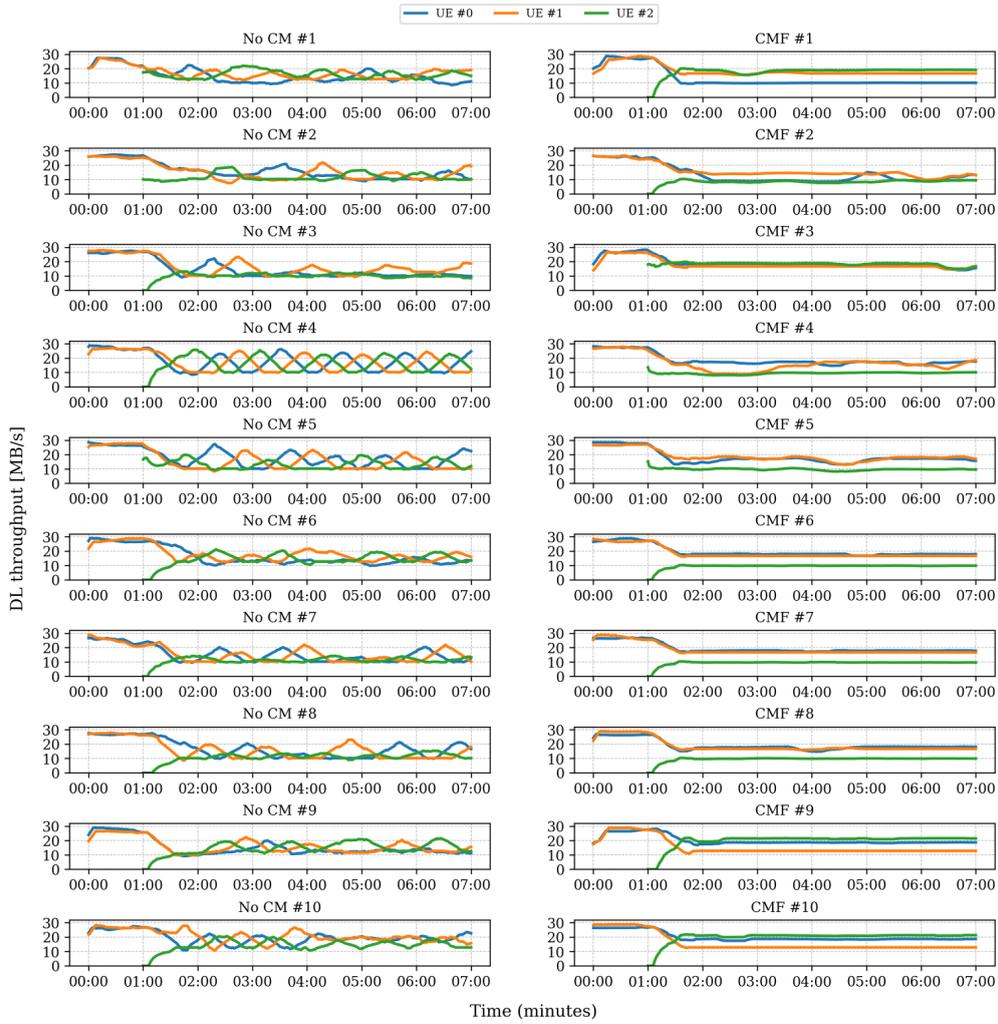

Fig. 4: DL throughput measurements captured during each of the test runs, with "No CM" and with "CMF" in place.

ACKNOWLEDGEMENTS

This work was supported by CCI xG Testbed. Visit CCI at: www.cyberinitiative.org. Cezary Adamczyk's contribution was funded by PUT. Adrian Kliks's involvement was supported by PUT and Rimedo Labs.